\documentclass[graybox]{svmult}

\usepackage{mathptmx}       
\usepackage{helvet}         
\usepackage{courier}        
\usepackage{type1cm}        
\usepackage{makeidx}         
\usepackage{graphicx}        
\usepackage{multicol}        
\usepackage[bottom]{footmisc}

\makeindex             

\begin{document}

\title*{The Transactional Interpretation\\
of Quantum Mechanics and\\
Quantum Nonlocality}
\author{John G. Cramer}
\institute{John G. Cramer \at Dept. of Physics, Univ. of Washington, Seattle, WA 98195-1560, \email{jcramer@uw.edu}}
%
%
\maketitle

\abstract{Quantum nonlocality is discussed as an aspect of the quantum formalism that is seriously in need of interpretation.  The Transactional Interpretation of quantum mechanics, which describes quantum processes as transactional ``handshakes" between retarded $\psi$ waves and advanced $\psi*$ waves, is discussed.  Examples of the use of the Transactional Interpretation in resolving quantum paradoxes and in understanding the counter-intuitive aspects of the formalism, particularly quantum nonlocality, are provided.}

\section{What \textit{is} an Interpretation of Quantum Mechanics?}
\sectionmark{What is an Interpretation of Quantum Mechanics?}
\label{sec:1}
Interpretations of quantum mechanics provide accounts of the meaning of the quantum formalism, guidance as to how to use the formalism to connect with nature and to make predictions on the outcome of experiments, and understanding of the counterintuitive aspects of the formalism.  The first interpretation of quantum mechanics was the Copenhagen Interpretation, developed by Werner Heisenberg and Niels Bohr in the late 1920s.  (See reference \cite{Cr86} for a complete description of and references for the Copenhagen Interpretation.)  It  quickly became the orthodox view of the meaning of the quantum formalism, and it is currently used in most quantum mechanics textbooks.  However, its ambiguities have  generated a large number of interpretational paradoxes associated with relativity conflicts, wave-particle duality, observer-dependent behavior, wave function collapse, and quantum nonlocality.

These problems generated by the Copenhagen Interpretation have led to a plethora of alternative ``interpretations", many of which are outlined in this book.  The reader is cautioned, however, to examine each of these interpretations carefully to determine if it really qualifies for the status of a \textit{full interpretation}, in that it deals with \textit{all} of the many interpretational problems inherent in the standard quantum formalism or raised by aspects of the Copenhagen Interpretation.

One would think that clever experimentalists could go into the quantum optics laboratory and determine which of these interpretations is correct by testing their experimental predictions.  However, this is not the case.  It is the \textit{formalism} of quantum mechanics that makes the testable experimental predictions, and all of the many interpretations are attempting to give meaning to that same formalism.  Thus, an interpretation could only be falsified if it was found to be inconsistent with the formalism of quantum mechanics, and otherwise the choice between interpretations becomes a matter of individual preference and philosophical aesthetics.

The interpretational problems of quantum nonlocality, which many would-be interpretations completely ignore, is a particularly difficult philosophical hurdle..  Many interpretational attempts instead focus on some particular problem, e.g., wave function collapse, to the exclusion of other interpretational  problems including nonlocality.  As we will see, the Transactional Interpretation is unique in providing a graphic picture of the mechanisms behind quantum nonlocality while dealing with all of the other interpretational problems as well. 

\section{Quantum Nonlocality}
\sectionmark{Quantum Nonlocality}
\label{sec:2}
Quantum mechanics, our standard theoretical model of the physical world at the smallest scales of energy and size, differs from the classical mechanics of Newton that preceded it in one very important way.  Newtonian systems are always \textit{local}.  If a Newtonian system breaks up, each of its parts has a definite and well-defined energy, momentum, and angular momentum, parceled out at breakup by the system while respecting conservation laws. After the component subsystems are separated, the properties of any subsystem are completely independent and do not depend on those of the other subsystems. 

On the other hand, quantum mechanics is \textit{nonlocal}, meaning that the component parts of a quantum system may continue to influence each other, even when they are well separated in space and out of speed-of-light contact.  This characteristic of standard quantum theory was first pointed out by Albert Einstein and his colleagues Boris Podolsky and Nathan Rosen (EPR) in 1935, in a critical paper\cite{Ei35}  in which they held up the discovered nonlocality as a devastating flaw that, it was claimed, demonstrated that the standard quantum formalism must be incomplete or wrong.  Einstein called nonlocality ``spooky actions at a distance".  Schr\"{o}dinger followed on the discovery of quantum nonlocality by showing in detail how the components of a multi-part quantum system must depend on each other, even when they are well separated\cite{Sc35}.

Beginning in 1972 with the pioneering experimental work of Stuart Freedman and John Clauser\cite{Fr72}, a series of quantum-optics EPR experiments testing Bell-inequality violations\cite{Be64}  and other aspects of entangled quantum systems were performed.  This body of experimental results can be taken as a demonstration that, like it or not, both quantum mechanics and the underlying reality it describes are intrinsically nonlocal.  Einstein's spooky actions-at-a-distance are really out there in the physical world, whether we understand and accept them or not.\\

How and why is quantum mechanics nonlocal?  Nonlocality comes from two seemingly conflicting aspects of the quantum formalism: (1) energy, momentum, and angular momentum, important properties of light and matter, are conserved in all quantum systems, in the sense that, in the absence of external forces and torques, their net values must remain unchanged as the system evolves, while (2) in the wave functions describing quantum systems, as required by Heisenberg's uncertainty principle\cite{He27}, the conserved quantities may be indefinite and unspecified and typically can span a large range of possible values. This non-specifity persists until a measurement is made that ``collapses" the wave function and fixes the measured quantities with specific values. These seemingly inconsistent requirements of (1) and (2) raise an important question: how can the wave functions describing the separated members of a system of particles, which may be light-years apart, have arbitrary and unspecified values for the conserved quantities and yet respect the conservation laws when the wave functions are collapsed? 

This paradox is accommodated in the formalism of quantum mechanics because the quantum wave functions of particles are \textit{entangled}, the term coined by Schr\"{o}dinger\cite{Sc35} to mean that even when the wave functions describe system parts that are spatially separated and out of light-speed contact, the separate wave functions continue to depend on each other and cannot be separately specified.  In particular, the conserved quantities in the system's parts (even though individually indefinite) must always add up to the values possessed by the overall quantum system before it separated into parts. 

How could this entanglement and preservation of conservation laws possibly be arranged by Nature?  The mathematics of quantum mechanics gives us no answers to this question, it only insists that the wave functions of separated parts of a quantum system do depend on each other.  Theorists prone to abstraction have found it convenient to abandon the three-dimensional universe and describe such quantum systems as residing in a many-dimensional Hilbert hyper-space in which the conserved variables form extra dimensions and in which the interconnections between particle wave functions are represented as allowed sub-regions of the overall hyper-space.  That has led to elegant mathematics, but it provides little assistance in visualizing what is really going on in the physical world.\\

Consider these questions:
\begin{itemize}
\item Is the quantum wave function a real object present in space-time?
\item What are the true roles of the observers and measurements in quantum processes?
\item What is wave function collapse?
\item How can quantum nonlocality be understood?
\item How can quantum nonlocality be visualized?
\item What are the underlying physical processes that make quantum nonlocality possible?
\end{itemize}

To our knowledge, the only interpretation that adequately answers these questions is the Transactional Interpretation of quantum mechanics\cite{Cr86,Cr88,Cr01}, which will be described in what follows.

\section{The One-Dimensional Transaction Model}
\sectionmark{The One-Dimensional Transaction Model}
\label{sec:3}

The starting point for the Transactional Interpretation of quantum mechanics is to view the ``normal" wave functions $\psi$ appearing in the wave-mechanics formalism as Wheeler-Feynman retarded waves, to view the complex-conjugated wave functions $\psi*$ as Wheeler-Feynman advanced waves, and to view the calculations in which they appear together as Wheeler-Feynman ``handshakes" between emitter and absorber\cite{Wh45}.

\begin{figure}
\center
  \includegraphics[width=6 cm]{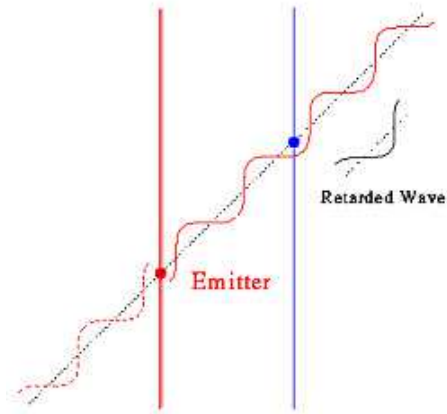}
\caption{(color online) Schematic of the emission stage of a transaction. An emiter produces a retarded wave (solid) toward an absorber and an advanced wave (dashed) in the other time direction.}
\label{fig:1}       
\end{figure}

While there are notable similarities between the Wheeler-Feynman time-symmetric approach to electrodynamics and this approach to the quantum formalism, there are also important differences.  In the classical electrodynamics of Wheeler-Feynman, it is the advanced-wave responses from all of the absorbers in the future universe, arriving together back at the emitter that cause it to radiate, lose energy, and recoil during emission.  There are no photons and there is no quantization of energy, and so there is no single future absorber that receives all of the energy and momentum that the emitter has transmitted.  Further, the emitter is responding to the full intensity of the superimposed advanced-wave fields from the future in a completely deterministic way, losing energy and gaining recoil momentum as a moving electric charge responding to external electric and magnetic fields.

In the domain of quantum mechanics these rules must be changed to reflect quantization and the probabilistic nature of quantum mechanics.  In the case of photon emission and absorption, an emitter emits a single photon, losing a quantum of energy and experiencing momentum recoil.  An absorber receives a single photon, gaining a quantum of energy $\hbar\omega$ and experiencing momentum recoil $\hbar k$.  The rest of the future universe does not explicitly participate in the process.  If the wave function  $\psi$ propagates for a significant distance before absorption, it becomes progressively weaker, much too weak to be consistent with the behavior of an electric charge simply responding to external fields as in classical Wheeler-Feynman electrodynamics.

\begin{figure}
\center
  \includegraphics[width=7 cm]{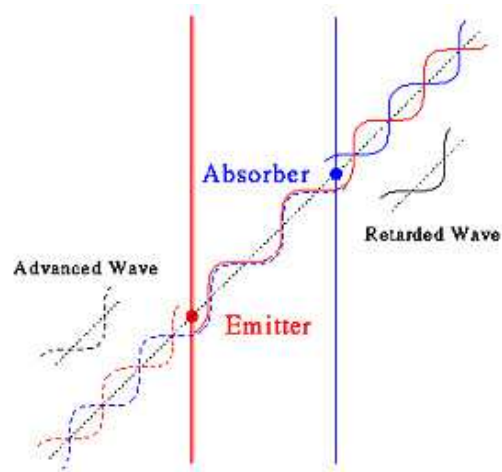}
\caption{(color online) Schematic of the confirmation stage of a transaction.  An absorber responds with an advance wave (dashed) back to the emmitter and a retarded wave (solid) going foward in time beyond the absorber.}
\label{fig:2}       
\end{figure}

As an intermediate conceptual step, it is useful to think about the quantum situation in a single space dimension $x$ and in one time dimension $t$, so that the attenuation of the wave function with distance can be put aside, for the moment.  This is a wave-on-a-string situation in which the light cone becomes a diagonal Minkowski line connecting emitter to absorber, as shown in Fig.\ref{fig:1}.  In the spirit of even-handed time symmetry, the emitter must simultaneously send out retarded wave function  $F_{1}(x, t) = \psi = A\exp[i(kx-\omega t)]$ and advanced wave function $G_{1}(x, t) =  \psi* = A\exp[-i(kx-\omega t)]$ in the two time and space directions, i.e., in both directions from the emitter along the Minkowski line.  The energy and momentum eigenvalues of $ F_{1}$ are $\hbar\omega$ and $\hbar k$, while the eigenvalues of $G_{1}$ are $-\hbar\omega$ and $-\hbar k$.  Therefore, the emission of the composite wave function $ F_{1}+ G_{1}$ involves no change in energy or momentum, i.e., it has no energy or momentum cost.   This is to be expected, since the emission process is time-symmetric, and time-symmetric fields should not produce any time-asymmetric loss of energy or momentum.

The absorber at some later time receives the retarded wave $ F_{1}$ and terminates it by producing a canceling wave $F_{2} = -A /exp[i(kx-\omega t)]$, as shown in Fig. \ref{fig:2}.  Because the absorber must respond in a time-symmetric way, it must also produce advance wave $G_{2} = -A \exp[-i(kx-\omega t)]$, which travels back along the Minkowski line until it reaches the emitter.  At the emitter it exactly cancels the advanced wave $G_{1}$ that the emitter had produced in the negative time direction.

\begin{figure}
\center
  \includegraphics[width=7 cm]{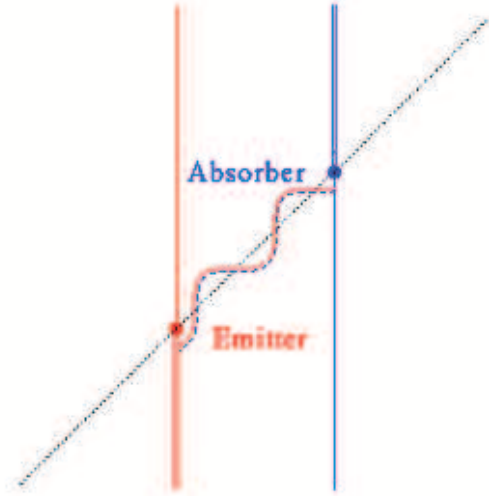}
\caption{(color online) Schematic of the completed transaction. Extra waves cancel, leaving an advanced-retarded ``handshake" that transfers energy $\hbar\omega$ and momentum $\hbar k$ from emitter to absorber.}
\label{fig:3}       
\end{figure}

The net result is that a superposition of $ F_{1}+G_{2}$ connects emitter with absorber, the emitting charge interacts with $G_{2}$ by losing energy $\hbar\omega$ and recoiling with momentum $-\hbar k$, and the absorbing charge interacts with $F_{1}$  by gaining energy $\hbar\omega$ and recoiling with momentum $\hbar k$.  Due to the cancellations beyond the interaction points, there is no wave function on the Minkowski diagonal before emission or after absorption.  A Wheeler-Feynman handshake, shown in Fig. \ref{fig:3}, has moved a quantum of energy  $\hbar\omega$ and momentum $\hbar k$  from emitter to absorber.  An observer, unaware of the time-symmetric processes involved, would say that a forward-going wave was emitted and subsequently absorbed.

\section{The Three-Dimensional Transaction Model}
\sectionmark{The Three-Dimensional Transaction Model}
\label{sec:4}

Now let us consider the more realistic situation of three spatial dimensions and one time dimension.  Now, assuming symmetric emission, the wave function $\psi$ spreads out in three dimensions like a bubble expanding from the central source location.  The wave function, attenuated by distance, can reach many potential absorbers, each of which can respond by producing an advanced wave function $\psi*$ that, also attenuated by distance, travels back to the emitter.  The emitter at the instant of emission can thus receive many advanced-wave ``echoes".  In this way, guided by the quantum formalism, attenuation and competition have been added to this picture.

We can think of the retarded waves from the emitter as offers to transmit energy or ``offer waves", the first step in a handshake process that may ultimately produce the emission of a photon.  Similarly, the advanced wave responses from potential absorbers can be thought of as ``confirmation waves", the second step in the handshake process to transfer a photon.   The advanced waves travel in the negative time direction and arrive back at the emission space-time location \textit{at the instant of emission}, each with a strength $\psi$ that reflects the attenuation of the offer wave in traveling forward from emitter to absorber multiplied by a strength $\psi*$ that reflects the attenuation of the confirmation wave in traveling back from absorber to emitter.

Therefore, the emitter receives an ``echo" of magnitude $\psi_{i}\psi_{i}*$ from the $i^{th}$ potential future absorber.  To proceed with the process, the emitter must ``choose" one (or none) of these offer-confirmation echoes as the initial basis for a photon-emission handshake or ``transaction", with the choice weighted in probability by the strength of each echo.  After the choice is made, there must be repeated emitter-absorber wave exchanges, until the strength of the space-time standing wave that thus develops is sufficient in strength to transfer a quantum of energy $\hbar\omega$ and momentum $\hbar k$ from the emitter to the absorber, completing the transaction.

As a criticiam of this transaction model, it might be argued that while the quantum wave function $\psi$ is a solution of the Schr\"{o}dinger wave equation, its complex conjugate $\psi*$ is not, and therefore the transaction model is inappropriately mixing solutions with non-solutions.  However, we observe that the Schr\"{o}dinger wave equation is inconsistent with Lorenz invariance and can be regarded as only the non-relativistic limit of the ``true" relativistic wave equation, i.e., the Klein-Gordon equation for bosons or the Dirac wave equation for fermions, both consistent with relativity.  Taking the non-relativistic limit of the  Klein-Gordon or Dirac wave equation produces \textit{two} wave equations, the Schr\"{o}dinger wave equation and its complex conjugate.  The wave function $\psi$ is a solution of the Schr\"{o}dinger wave equation, while $\psi*$ is a solution of the complex conjugate of Schr\"{o}dinger wave equation, and so both are equally valid solutions.  The quantum version of the electromagnetic wave equation, which is relativistically invariant and is appropriate for describing the emission and absorption of photons, has both advanced and retarded solutions.

We note here that the sequence of stages in the emitter-absorber transaction presented here employs the semantic device of ``pseudo-time", describing a process between emitter and absorber extending across lightlike or timelike intervals of space-time as if it occurred in a time sequence external to the process. This is only a pedagogical convention for the purposes of description. The process itself is atemporal, and the only observables come from the superposition of all of the steps that form the final transaction.

This is the transaction model by which the Transactional Interpretation describes the elements of the wave-mechanics formalism and accounts for quantum mechanical processes.  The wave functions $\psi$ of the wave-mechanics formalism are the offer waves.  In some sense they are real waves traveling through space, but in another sense they are not real because they represent only a mathematical encoding of the \textit{possibility} of a quantum process.  The transaction that forms after the emitter-absorber offer-confirmation exchange process goes to completion is the real object, what we would call the ``particle" that has been transferred from emitter to absorber.  In that sense, the real objects in our universe are waves, while particles are an illusion created by the boundary conditions that must be observed at the vertices of the wave-exchange transactions.

What happens to the offer and confirmation waves that do not result in the formation of a transaction?   Since the formation of a transaction produces all of the observable effects, such waves are ephemeral, in that they produce no observable effects, and their presence or absence has no physical consequences.  However,  in explaining seemingly paradoxical quantum phenomena such as interaction-free measurements\cite{El93,Cr06}, such waves can be viewed as ``feeling out" components of the system even when no transaction forms.

The transactional model not only provides a description of the process that underlies the calculation of a quantum mechanical matrix element, but it also explains and justifies Born's probability rule\cite{Bo26}.  In particular, it explains why a quantum event described by a wave function $\psi$  has a probability of occurrence given by $\psi\psi*$.  In the transaction model, the quantities $\psi\psi*$  are the strengths of the advanced-wave echoes arriving back at the site of emission at the instant of emission.  The ``lightning strike" of a transaction formation depends probabilistically on the strengths of these echoes.

The Born probability rule is an assumption of the Copenhagen Interpretation, asserted axiomatically without justification as one of the tenets of the interpretation.  On the other hand, the Born probability rule follows naturally from the transactional account of  the Transactional Interpretation and does not need to be added as a separate assumption.  In that sense, the Transactional Interpretation is superior to the Copenhagen Interpretation because it is more philosophically ``economical", requiring fewer independent assumptions.\\

There is one more element of the transaction model, \textit{hierarchy}, which needs to be added in order to avoid transactional inconsistencies pointed out by Maudlin\cite{Ma96}.  All advanced-wave echoes are not equal.  Those propagating back to the emitter from small space-time separation intervals rank higher in the selection hierarchy than those propagating back to the emitter from large space-time separation intervals.  The emitter's probabilistic decision to select or not select an echo propagating back from a small space-time interval must occur ``before" any echoes from larger space-time intervals are considered and their transaction allowed to form. 

This hierarchy of transaction formation has interesting implications for time itself in quantum processes.  In some sense, the entire future of the universe is reflected in the formation of each transaction, with the echoes from time-distant future events allowed the possibility of forming transactions only after the echoes from near future absorbers have been weighed and rejected.
To make another analogy, the emergence of the unique present from the future of multiple possibilities, in the view of the Transactional Interpretation, is rather like the progressive formation of frost crystals on a cold windowpane, first nearby and then extending further out.  As the frost pattern expands, there is no clear freeze-line, but rather a moving boundary, with fingers of frost reaching out well beyond the general trend, until ultimately the whole window pane is frozen into a fixed pattern.  In the same way, the emergence of the present involves a lace-work of connections with the future and the past, insuring that the conservation laws are respected and the balances of energy and momentum are preserved.

\section{The Transactional Interpretation of Quantum Mechanics}
\sectionmark{The Transactional Interpretation of Quantum Mechanics}
\label{sec:5}

The Transactional Interpretation of quantum mechanics\cite{Cr86, Cr88,Cr01}, inspired by the structure of the quantum wave mechanics formalism itself, views each quantum event as a Wheeler-Feynman ``handshake" or ``transaction" process extending across space-time that involves the exchange of advanced and retarded quantum wave functions to enforce the conservation of certain quantities (energy, momentum, angular momentum, etc.).  It asserts that each quantum transition forms in four stages: (1) emission, (2) response, (3) stochastic choice, and (4) repetition to completion.

The first stage of a quantum event is the emission of an ``offer wave" by the ``source", which is the object supplying the quantities transferred.  The offer wave is the time-dependent retarded quantum wave function $\psi$, as used in standard quantum mechanics.  It spreads through space-time until it encounters the ``absorber", the object receiving the conserved quantities.

The second stage of a quantum event is the response to the offer wave by any potential absorber (there may be many in a given event). Such an absorber produces an advanced ``confirmation wave" $\psi*$, the complex conjugate of the quantum offer wave function $\psi$. The confirmation wave travels in the reverse time direction and arrives back to the source at precisely the instant of emission with an amplitude of $\psi\psi*$.

The third stage of a quantum event is the stochastic choice that the source exercises in selecting one of the many received confirmations. The strengths $\psi\psi*$ of the advanced-wave ``echoes" determine which transaction forms in a linear probabilistic way.

The final stage of a quantum event is the repetition to completion of this process by the source and selected absorber, reinforcing the selected transaction repeatedly until the conserved quantities are transferred and the potential quantum event becomes real.\\

Here we summarize the principal elements of the Transactional Interpretation, structured in order to contrast it with the Copenhagen Interpretation:

\begin{itemize}

\item The fundamental quantum mechanical interaction is taken to be the transaction.  The state vector $\psi$ of the quantum mechanical formalism is a physical wave with spatial extent and is identical with the initial ``offer wave" of the transaction.  The complex conjugate of the state vector $\psi*$ is also a physical wave and is identical with the subsequent ``confirmation wave" of the transaction.  The particle (photon, electron, etc.) and the collapsed state vector are identical with the completed transaction.  The transaction may involve a single emitter and absorber and two vertices or multiple emitters and absorbers and many vertices, but is only complete when appropriate quantum boundary conditions are satisfied at all vertices, i.e., loci of emission and absorption.  Particles transferred have no separate identity independent from the satisfaction of the boundary conditions at the vertices.

\item The correspondence of ``knowledge of the system" with the state vector $\psi$ is a fortuitous but deceptive consequence of the transaction, in that such knowledge must necessarily follow and describe the transaction.

\item Heisenberg's Uncertainty Principle\cite{He27} is a consequence of the fact that a transaction in going to completion is able to project out and localize only one of a pair of conjugate variables (e.g., position or momentum) from the offer wave, and in the process it delocalizes the other member of the pair, as required by the mathematics of Fourier analysis.  Thus, the Uncertainty Principle is a consequence of the transactional model and is not a separate assumption.

\item Born's Probability Rule\cite{Bo26} is a consequence of the fact that the magnitude of the ``echo" received by the emitter, which initiates a transaction in a linear probabilistic way, has strength $P = \psi\psi*$.  Thus, Born's Probability Rule is a consequence of the transactional model and is not a separate assumption of the interpretation.

\item All physical processes have equal status, with the observer, intelligent or otherwise, given no special status.  Measurement and measuring apparatus have no special status, except that they happen to be processes that connect and provide information to observers.

\item Bohr's ``wholeness" of measurement and measured system exists, but is not related to any special character of measurements but rather to the connection between emitter and absorber through the transaction.

\item Bohr's ``complementarity" between conjugate variables exists, but like the uncertainty principle is just a manifestation of the requirement that a given transaction going to completion can project out only one of a pair of conjugate variables, as required by the mathematics of Fourier analysis.

\item Resort to the positivism of ``don't-ask-don't-tell" is unnecessary and undesirable.  A distinction is made between observable and inferred quantities.  The former are firm predictions of the overall theory and may be subjected to experimental verification.  The latter, particularly those that are complex quantities, are not verifiable and are useful only for visualization, interpretational, and pedagogical purposes.  It is assumed that both kinds of quantities must obey conservation laws, macroscopic causality conditions, relativistic invariance, etc. 

\end{itemize}

In summary, the Transactional Interpretation explains the origin of the major elements of the Copenhagen Interpretation while avoiding their paradoxical implications.  It drops the positivism of the Copenhagen Interpretation as unnecessary, because the positivist curtain is no longer needed to hide the nonlocal backstage machinery.

It should also be pointed out that giving some level of objective reality to the state vector colors all of the other elements of the interpretation.  Although in the Transactional Interpretation, the uncertainty principle and the statistical interpretation are formally the same as in the Copenhagen Interpretation, their philosophical implications, about which so much has been written from the Copenhagen viewpoint, may be rather different.

The Transactional Interpretation offers the possibility of resolving \textit{all} of the many interpretational paradoxes that quantum mechanics has accumulated over the years.  Many of these are analyzed in reference \cite{Cr86}, the publication in which the Transactional Interpretation was introduced.  Here we will not attempt to deal with all of the paradoxes.  We will instead focus on the interpretational problems associated with quantum nonlocality and entanglement.

\section{The Transactional Interpretation and Nonlocality}
\sectionmark{The Transactional Interpretation and Nonlocality}
\label{sec:6}

As we discussed in Section \ref{sec:2}, quantum nonlocality is one of the principal counter-intuitive aspects of quantum mechanics.  Einstein's ``spooky action-at-a-distance" is a real feature of quantum mechanics, but the quantum formalism and the orthodox Copenhagen Interpretation provide little assistance in understanding nonlocality or in visualizing what is going on in a nonlocal process.  The Transactional Interpretation provides the tools for doing this.

Perhaps the first example of a nonlocality paradox is the Einstein's bubble paradox, proposed by Albert Einstein at the 5th Solvay Conference in 1927\cite{Ja66}.  A source emits a single photon isotropically, so that there is no preferred emission direction.  According the the quantum formalism, this should produce a spherical wave function $\psi$ that expands like an inflating bubble centered on the source.  At some later time, the photon is detected, and according to the quantum formalism, the bubble visualized as the wave function should ``pop", disappearing instantaneously from all locations except the position of the detector.  Einstein asked how the parts of the wave function away from the detector could ``know" that they should disappear, and how it could be arranged that only a single photon was detected?

\begin{figure}
\center
  \includegraphics[width=9 cm]{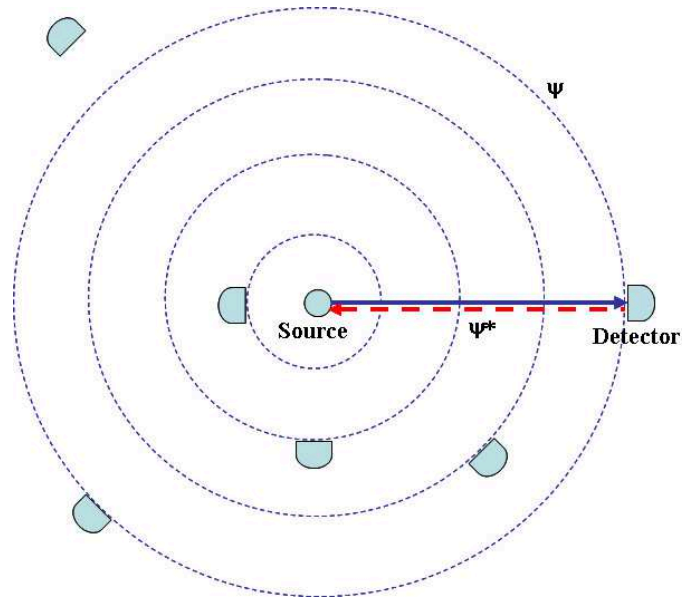}
\caption{(color online) Schematic of the transaction involved in the Einstein's bubble paradox.  The offer wave $\psi$ forms a spherical wave front, reaching the detector on the right and causing it to return a confirmation wave $\psi*$, so that a transaction forms and a photon of energy is transferred. Other detectors also return confirmation waves, but the source has randomly selected the detector on the right for a transaction.}
\label{fig:4}       
\end{figure}

At the 5th Solvay Conference, Werner Heisenberg\cite{Ja66} dismissed Einstein's bubble paradox by asserting that the wave function is not a real object moving through space, as Einstein had implicitly assumed, but instead is a mathematical representation of the knowledge of some observer who is watching the process.  Until  detection, the observer knows nothing about the location of the emitted photon, so the wave function must be spherical, distributed over the $4\pi$ solid angle to represent his ignorance.  However, after detection the location of the photon is known to the observer, so the wave function ``collapses" and is localized at the detector.  One photon is detected because only one photon was emitted.

The Transactional Interpretation provides an alternative explanation, one that permits the wave function to be, in some sense, a real object moving through space.  This is illustrated in Fig. \ref{fig:4}.  The offer wave $\psi$ from the source indeed spreads out as a spherical wave front and eventually encounters the detector on the right.  The detector responds by returning to the source a confirmation wave $\psi*$.  Other detectors (i.e., potential absorbers) also return confirmation waves, but the source randomly, weighted by the $\psi\psi*$ echoes from the potential absorbers,  selects the detector on the right to form a transaction.  The transaction forms between source and detector, and one $\hbar\omega$ photon's worth of energy is transferred from the source to the detector.  The formation of this particular transaction, satisfying the source boundary condition that only one photon is emitted, prevents the formation of any other transaction to another possible photon absorber, so only one photon is detected.  Further, the wave function bubble never pops, as Einstein had suggested.  The unblocked parts of the offer wave keep moving outward and encountering more distant absorbers, which participate in the competition for transaction formation.

One might ask about the ``left over" offer and confirmation waves that do not participate in the formation of a transcation.  However, we note that the first two stages (offer and confirmation) of transaction formation map directly into classical Wheeler-Feynman time-symmetric electrodynamics\cite{Wh45}, in which the advanced and retarded waves before emission and after absorption largely cancel out, leaving little in the way of residue.

This is an illustration of a simple two-vertex transaction in which the transfer of a single photon is implemented nonlocally.  It avoids Heisenberg's peculiar assertion that the mathematical solution to a simple 2nd order differential equation involving momentum, energy, time, and space is somehow a map of the mind, deductions, and knowledge of a hypothetical observer.\\

As another example of nonlocality in action, consider the Freedman-Clauser experiment\cite{Fr72}.  An atomic transition source produces a pair of polarization-entangled photons.  The three-level atomic transitions that produce the two photons have a net orbital angular momentum change of $L=0$ and even parity, so, if the photons are emitted back-to-back, they must both be in the same state of circular polarization or linear polarization.  Measurements on the photons with linear polarimeters in each arm of the experiment show that when the planes of the polarimeters are aligned, independent of the direction of alignment, the two polarimeters always measure HH or VV for the two linear polarization states.

When the plane of one polarimeter is rotated by an angle $\theta$ with respect to the other plane, some opposite-correlation HV and VH events creep in, and these grow as $1-\cos^{2}(\theta)$, which for small values of $\theta$ is proportional to $\theta^2$.  This polarization correlation behavior produces a dramatic violation of the Bell inequalities\cite{Be64}, which for local hidden variable alternatives to standard quantum mechanics require a \textit{linear} growth in HV and VH with $\theta$.   The implication of the Bell-inequality violations is that quantum nonlocality is required to explain the observed quadratic polarization correlations.

\begin{figure}
\center
  \includegraphics[width=11 cm]{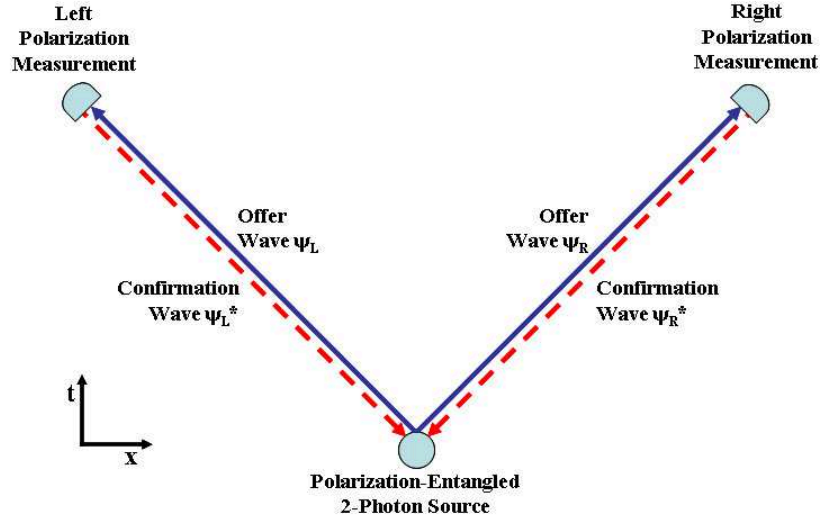}
\caption{(color online) Space-time schematic of a nonlocal ``V" transaction for visualizing the polarization-entangled Freedman-Clauser EPR experiment.  Offer waves $\psi_{L}$ and $\psi_{R}$ (blue/solid) move from source to detectors, and in response, confirmation waves $\psi_{L}*$ and $\psi_{R}*$ (red/dashed)  move from detectors to source.  The three-vertex transaction can form only if angular momentum is conserved by having correlated and consistent measured polarizations for both detected photons.}
\label{fig:5}       
\end{figure}

How are the nonlocality-based polarization correlations of the Freedman-Clauser experiment possible?  The Transactional Interpretation provides a clear answer, which is illustrated in Fig. \ref{fig:5}.  The source of the polarization-entangled photons seeks to emit a photon pair by sending out offer waves  $\psi_{L}$ and $\psi_{R}$ to the left and right detectors.  The detectors respond by returning confirmation waves $\psi_{L}*$ and $\psi_{R}*$ back to the source.  A completed three-vertex transaction can form from these, however, only if the two potential detections are compatible with the conservation of angular momentum at the source.  This requirement produces the observed polarization correlations.

There are also a large number of more complicated experiments that demonstrate the peculiarities of quantum nonlocality and entanglement in other ways, and which involve transactions with more than three vertices.  One such example is the interaction-free measurement experiment of Elitzur and Vaidmann\cite{El93}, which has been analyzed with the Transactional Interpretation\cite{Cr06}.  In all such cased, the Transactional Interpretation provides a way of visualizing multi-vertex nonlocal processes that otherwise seem strange and counter-intuitive.

\section{ Do Quantum Wave Functions Exist in Real Three-Dimensional Space or Only in Hilbert Space?}
\sectionmark{ Do Quantum Wave Functions Exist in Real 3D Space?}
\label{sec:7}

In classical wave mechanics, propagating waves, e.g. light or sound waves, are viewed as existing in and propagating through normal three-dimensional space.  However, early in the development of the formalism of quantum mechanics it was realized that there was a problem with treating the quantum wave functions of multi-particle systems in the same way.  Because of conservation laws and entanglement, the uncollapsed wave function of each particle in such a system was not only a function of its own space and momentum coordinates and other variables (e.g. spin and angular momentum), but might also be dependent on the equivalent coordinates of the other particles in the system.  For example, the momentum magnitude and direction of each particle of a quantum system might be unspecified and allowed to take on any value over a wide range, but their momenta must be correlated so that the overall momentum of the system can have a well defined momentum value.

Therefore, it was concluded that Hilbert space provided a general way of describing quantum systems and that in multi-particle systems, a quantum mechanical wave function could not exist in simple three-dimensional space, but must instead reside in a higher Hilbert space of many more dimensions, with a dimension for each relevant variable.    The wave function of a ``free" independent non-entangled particle in such a space simply ``traces over" the extra Hilbert space dimensions, allowing the extra variables to take on any value because there is no dependence on them.  In such a Hilbert space the inter-dependences of multi-particle systems could be described, conservation laws could be defined as ``allowed regions" that the wave functions could occupy, and powerful mathematical operations appropriate to higher dimensional spaces could be applied to the quantum formalism.  The assertion that quantum wave functions cannot be considered to exist in normal space and must be viewed as existing only in an abstract higher-dimensional space, of course, creates a severe roadblock for any attempt to visualize quantum processes.  (We note that Ruth Kastner's ``Possibilist Transactional Interpretation"\cite{Ka12,Ka13} adopts this point of view and treats quantum wave functions as being real objects only in an abstract multidimensional Hilbert space, from which transactions emerge in real space.  The possibilist approach is not incorrect,  but we consider it to be unnecessarily abstract.)

The ``standard" Transactional Interpretation, with its insights into the mechanism behind wave function collapse through multi-vertex transaction formation, provides a new view of the situation that make the retreat to Hilbert space unnecessary.  The offer wave for each particle can be considered as the wave function of a free particle, initially free of the constraints of conservation laws and indepent of the characteristics of other particles,  and can be viewed as existing in normal three dimensional space.  The connections between an ensemble of such free particles is only established when the multi-vertex transaction forms.  The application of conservation laws and the influence of the variables of the other particles of the system comes not in the initial offer wave stage of the process but in the final formation of the transaction  The transaction ``knits together" the various otherwise independent particle wave functions that can span a wide range of possible parameter values into an interaction, and only those wave function sub-components that correlate to satisfy the conservation law boundary conditions can participate in the final multi-vertex transaction formation.  The ``allowed zones" of Hilbert space arise from the action of transaction formation, not from constraints on the initial offer waves, i.e., particle wave functions.  Hilbert space is the map, not the territory.

Thus, the assertion that the quantum wave functions of individual particles in a multi-particle quantum system cannot exist in ordinary three-dimensional space might be a misinterpretation of the role of Hilbert space, the application of conservation laws, and the origins of entanglement.  Offer waves are somewhat ephemeral three-dimensional space objects, but only those components of the offer wave that satisfy conservation laws and entanglement criteria are permitted to be projected out in the final transaction, which also exists in three-dimensional space.

Another interesting question, relevant to the current need for a yet-unknown theory of quantum gravity, is whether the Transactional Interpretation would be consistent with the existence of a ``universal" quantum wave function that could describe the state of the entire universe.  The Copenhagen Interpretation, with its focus on observers, has a severe problem with a universal wave function that would be interpreted as a description of observer knowledge and would require an observer outside the universe to collapse it.  The Transactional Interpretation, which is independent of observers and observer knowledge, has no such problems.  Further, it is relativistically invariant, and therefore could, in principle, be extended to a theory of quantum gravity, should one that used wave functions emerge from the current theoretical effort.

The two examples of nonlocality-based \textit{gedankenexperiments} presented here provide only a sample of the power of the Transactional Interpretation in analyzing the complex and counter-intuitive experimental results that seem to be emerging from experimental quantum optics at an exponentially increasing rate.  The transactional analysis of interaction-free measurments\cite{El93,Cr06} is another example of the power of the method.  However, as the experiments become more complex, the analysis inevitably becomes more elaborte and difficult to follow.  Therefore, for the purposes of the present discussion, we will confine ourselves to the above two analysis examples.

\section{Conclusion}
\sectionmark{Conclusion}
\label{sec:8}

Is free will possible in such a system? It is our view that it is. Freedom of choice does not include the freedom to choose to violate physical laws.  The transactional handshakes between present and future are acting to enforce physical laws, and they restrict the choices between future possibilities only to that extent.

By analogy, when you present a debit card to purchase groceries, there is a nearly-instantaneous electronic transaction between the cash register and the bank that deducts the purchase cost from your bank account and insures that you have sufficient funds for the purchase.  The bank transaction enforces some ``law of conservation of money" as applied to your finances.  But the transaction is not deterministic, and in particular it does not determine what you buy, only that you can afford what you have bought.  This is similar to what goes on in a photon emission-absorption transaction, the transaction ensuring that precisely one photon-worth of energy and momentum is deducted from one system and added to another system.\\

We conclude that the Transactional Interpretation does not require (but \textit{is} consistent with) a deterministic block universe. It does, however, imply that the emergence of present reality from future possibility is a far more hierarchical and complex process than we have previously been able to imagine.

We have seen that the Transactional Interpretation of quantum mechanics provides the tools for understanding the many counterintuitive aspects of the quantum formalism and for visualizing nonlocal quantum processes.  Further, the transaction model is ``visible" in the quantum formalism itself, once one associates the wave function $\psi$ with an offer, the conjugated wave function $\psi*$ with a confirmation, and quantum matrix elements with completed transactions.

To our knowledge, the Transactional Interpretation of quantum mechanics is the only interpretation that adequately answers the questions arising from quantum nonlocality and entanglement and also deals with all of the other interpretational problems of the quantum formalism.

\begin{acknowledgement}
The author would like to thank the many people who, over the years, have made comments and raised interesting questions that can be addressed with the Transactional Interpretation.   These include the late Sir Rudolph Peierls, the late John Wheeler, John Clauser, Rudolph M\"{o}ssbauer, Avshalom Elitzur, Ruth Kastner, Gerald Miller, Vince  Feng, James Woodward, Heidi Fearn, and Nick Herbert.
\end{acknowledgement}

\end{document}